\documentclass[1p]{elsarticle}
\biboptions{sort&compress}
\journal{Nuclear Physics A}
\usepackage{graphicx}
\usepackage{amsmath}
\usepackage{amssymb}
\usepackage{mathtools}
\usepackage{liealg}
\usepackage{wigner}

\DeclareRobustCommand{\underbracketnote}[2]{\underbracket[1pt]{#1}_{\text{#2}}}

\DeclareRobustCommand{\underbracevalue}[2]{\underbrace{#1}_{\mathclap{#2}}}
\DeclareRobustCommand{\underbracenovalue}[1]{\underbrace{#1}}

\newlength{\underarrowlength}
\DeclareRobustCommand{\setunderarrowlength}[1]{\settowidth{\underarrowlength}{\ensuremath{#1}}\addtolength{\underarrowlength}{-0.5em}}
\DeclareRobustCommand{\underarrow}[2]{\setunderarrowlength{#2}
\underset{\mathclap{\csname x#1arrow\endcsname{\rule{\underarrowlength}{0pt}}}}{#2}}
\DeclareRobustCommand{\underarrownote}[3]{\setunderarrowlength{#2}
\underset{\mathclap{\csname x#1arrow\endcsname[\mathclap{#3}]{\rule{\underarrowlength}{0pt}}}}{#2}}

\newlength{\widetildedaggerheight}
\DeclareRobustCommand{\widetildedagger}[1]{
\settoheight{\widetildedaggerheight}{\ensuremath{#1}}
\widetilde{#1}\rule{0pt}{\widetildedaggerheight}^\dagger
}

\def\C(#1){C_{#1}}
\def\Ct(#1){\tilde{C}_{#1}}
\def\Cd(#1){C_{#1}^\dagger}

\def\A(#1#2#3){A_{{#1}{#2}}^{#3}}
\def\At(#1#2#3){\tilde{A}_{{#1}{#2}}^{#3}}
\def\Ad(#1#2#3){A_{{#1}{#2}}^{#3\,\dagger}}

\def\S(#1){S^{#1}}
\def\St(#1){\tilde{S}^{#1}}
\def\Sd(#1){S^{\dagger\,#1}}

\def\T(#1#2#3){T_{{#1}{#2}}^{#3}}
\def\TBOCCCC(#1#2#3#4#5#6#7){[(\Cd({#1})\times\Cd({#2}))^{#3}\times(\Ct({#4})\times\Ct({#5}))^{#6}]^{#7}}
\def\TBOAA(#1#2#3#4#5#6#7){(\Ad(#1#2#3)\times\At(#4#5#6))^{#7}}

\def\E(#1){E^{#1}}
\def\F(#1){F^{#1}}
\def\G(#1){G^{#1}}
\def\H(#1){H^{#1}}
\def\I(#1){I^{#1}}

\def\Et(#1){\tilde{E}^{#1}}
\def\Ft(#1){\tilde{F}^{#1}}
\def\Gt(#1){\tilde{G}^{#1}}
\def\Ht(#1){\tilde{H}^{#1}}
\def\It(#1){\tilde{I}^{#1}}

\def\Ed(#1){E^{#1\,\dagger}}
\def\Fd(#1){F^{#1\,\dagger}}
\def\Gd(#1){G^{#1\,\dagger}}
\def\Hd(#1){H^{#1\,\dagger}}
\def\Id(#1){I^{#1\,\dagger}}

\def\P(#1#2#3){\bar{\mathbb{P}}(#1#2;#3)}
\def\PP#1#2{\mathbb{P}(#1\leftrightarrow#2)}

\def\lclusterC(#1){#1}
\def\lclusterA(#1#2#3){(#1#2)^{#3}}
\def\lclusterAC(#1#2#3#4#5){(#1#2)^{#3}#4^{#5}}
\def\lclusterAA(#1#2#3#4#5#6#7){(#1#2)^{#3}(#4#5)^{#6\,#7}}
\def\lclusterAAC(#1#2#3#4#5#6#7#8#9){(#1#2)^{#3}(#4#5)^{#6\,#7}#8^#9}

\let\pclusterC\lclusterC  
\let\pclusterA\lclusterA  
\let\pclusterAC\lclusterAC 
\let\pclusterAA\lclusterAA

\def\dclusterC(#1){\Cd(#1)}
\def\dclusterA(#1#2#3){\Ad(#1#2#3)}
\def\dclusterAC(#1#2#3#4#5){(\Ad(#1#2#3)\times\Cd(#4))^{#5}}
\def\dclusterAA(#1#2#3#4#5#6#7){(\Ad(#1#2#3)\times\Ad(#4#5#6))^#7}
\def\dclusterAAC(#1#2#3#4#5#6#7#8#9){[(\Ad(#1#2#3)\times\Ad(#4#5#6))^#7\times\Cd(#8)]^#9}

\def\OLAP{O}
\def\OLAPsymbol#1#2{\OLAP_{#2}^{(#1)}}
\def\OLAPframe#1#2#3{\OLAPsymbol{#1}{#2}[#3]}
\def\OLAPzero(#1){\OLAP_{#1}^{(0)}}
\def\OLAPone(#1)[#2|#3]{\OLAPframe{1}{#1}{\lclusterC(#2) \vert \lclusterC(#3)}}
\def\OLAPtwo(#1)[#2|#3]{\OLAPframe{2}{#1}{\lclusterA(#2) \vert \lclusterA(#3)}}
\def\OLAPthree(#1)[#2|#3]{\OLAPframe{3}{#1}{\lclusterAC(#2) \vert \lclusterAC(#3)}}
\def\OLAPfour(#1)[#2|#3]{\OLAPframe{4}{#1}{\lclusterAA(#2) \vert \lclusterAA(#3)}}
\def\OLAPfive(#1)[#2|#3]{\OLAPframe{5}{#1}{\lclusterAAC(#2) \vert \lclusterAAC(#3)}}

\def\OBME{T}
\def\OBMEsymbol#1#2{\OBME_{#2}^{(#1)}}
\def\OBMEframe#1#2#3{\OBMEsymbol{#1}{#2}[#3]}
\def\OBMEzero(#1)[#2]{\OBMEframe{0}{#1}{\lclusterA(#2)}}
\def\OBMEone(#1)[#2|#3|#4]{\OBMEframe{1}{#1}{\lclusterC(#2) \vert \lclusterA(#3) \vert \lclusterC(#4)}}
\def\OBMEtwo(#1)[#2|#3|#4]{\OBMEframe{2}{#1}{\lclusterA(#2) \vert \lclusterA(#3) \vert \lclusterA(#4)}}
\def\OBMEthree(#1)[#2|#3|#4]{\OBMEframe{3}{#1}{\lclusterAC(#2) \vert \lclusterA(#3) \vert \lclusterAC(#4)}}
\def\OBMEfour(#1)[#2|#3|#4]{\OBMEframe{4}{#1}{\lclusterAA(#2) \vert \lclusterA(#3) \vert \lclusterAA(#4)}}
\def\OBMEfive(#1)[#2|#3|#4]{\OBMEframe{5}{#1}{\lclusterAAC(#2) \vert \lclusterA(#3) \vert \lclusterAAC(#4)}}

\def\TBME{W}
\def\TBMEsymbol#1#2{\TBME_{#2}^{(#1)}}
\def\TBMEframe#1#2#3{\TBMEsymbol{#1}{#2}[#3]}
\def\TBMEzero(#1)[#2]{\TBMEframe{0}{#1}{\lclusterAA(#2)}}
\def\TBMEone(#1)[#2|#3|#4]{\TBMEframe{1}{#1}{\lclusterC(#2) \vert \lclusterAA(#3) \vert \lclusterC(#4)}}
\def\TBMEtwo(#1)[#2|#3|#4]{\TBMEframe{2}{#1}{\lclusterA(#2) \vert \lclusterAA(#3) \vert \lclusterA(#4)}}
\def\TBMEthree(#1)[#2|#3|#4]{\TBMEframe{3}{#1}{\lclusterAC(#2) \vert \lclusterAA(#3) \vert \lclusterAC(#4)}}
\def\TBMEfour(#1)[#2|#3|#4]{\TBMEframe{4}{#1}{\lclusterAA(#2) \vert \lclusterAA(#3) \vert \lclusterAA(#4)}}
\def\TBMEfive(#1)[#2|#3|#4]{\TBMEframe{5}{#1}{\lclusterAAC(#2) \vert \lclusterAA(#3) \vert \lclusterAAC(#4)}}

\DeclareRobustCommand{\calP}{\mathcal{P}}
\DeclareRobustCommand{\calNt}{\tilde{\mathcal{N}}}

\DeclareRobustCommand{\deltasqrt}[1]{(1+\delta_{#1})^{1/2}}
\DeclareRobustCommand{\deltasqrtinv}[1]{(1+\delta_{#1})^{-1/2}}

\DeclareRobustCommand{\sumpart}[2]{\sum_{{\substack{#1\in\calP(#2)}}}}
\DeclareRobustCommand{\prodsp}[1]{\prod_{#1}}

\DeclareRobustCommand{\vev}[1]{\tme{0}{#1}{0}}
\DeclareRobustCommand{\vket}{\tket{0}}

\begin{document}

\title{\boldmath Recursive calculation of matrix elements for the
generalized seniority shell model}

\author{F. Q. Luo}
\author{M. A. Caprio\corref{cor1}}
\address{Department of Physics, University of Notre Dame,
Notre Dame, Indiana 46556-5670, USA}
\cortext[cor1]{Corresponding author.}

\date{\today}

\begin{abstract}
A recursive calculational scheme is developed for matrix elements in
the generalized seniority scheme for the nuclear shell model.
Recurrence relations are derived which permit straightforward and
efficient computation of matrix elements of one-body and two-body
operators and basis state overlaps.
\end{abstract}


\makeatletter
\def\keyword{%
 \def\sep{\unskip; }%
 \def\MSC{\@ifnextchar[{\@MSC}{\@MSC[2000]}}
  \def\@MSC[##1]{\par\leavevmode\hbox {\it ##1~MSC:\space}}%
  \def\PACS{\par\leavevmode\hbox {\it PACS:\space}}%
  \def\JEL{\par\leavevmode\hbox {\it JEL:\space}}%
  \global\setbox\keybox=\vbox\bgroup\hsize=\textwidth
  \normalsize\normalfont\def\baselinestretch{1}
  \parskip\z@
  \noindent\textit{Keywords: }
  \raggedright 
  \ignorespaces}
\makeatother  

\begin{keyword}
Generalized seniority \sep Shell model
\PACS 21.60.Cs
\end{keyword}
\maketitle


\section{Introduction}
\label{sec-intro}

The generalized
seniority~\cite{talmi1971:shell-seniority,shlomo1972:gen-seniority} or
broken pair~\cite{gambhir1969:bpm,allaart1988:bpm} approximation
provides a truncation scheme for the nuclear shell model, based on the
dominance of like-nucleon pairing effects in semimagic (or nearly
semimagic) nuclei.  The fundamental premise is that there exists an
energetically favored collective $S$ pair, which can be constructed
from like nucleons coupled to angular momentum zero, and that the
ground state of an even-even nucleus can be well approximated by a
condensate of such collective pairs.  Low-lying excited states may
then be obtained by breaking one or more $S$ pairs.  The resulting
states are classified by the generalized seniority $v$, defined as the
number of valence nucleons not participating in a collective $S$ pair.
The generalized seniority approach effectively reduces the
dimensionality of the $n$-particle valence shell model to that of a
$v$-particle shell model problem.  In practice, such generalized
seniority shell model calculations have been carried out for $v\leq4$
(\textit{e.g.},
Refs.~\cite{bonsignori1978:tda-sn,scholten1983:gssm-n82,bonsignori1985:bpm-sn,engel1989:gssm-double-beta,monnoye2002:gssm-ni}).

Moreover, the generalized seniority scheme provides a microscopic
foundation for the phenomenologically successful interacting boson
model (IBM)~\cite{iachello1987:ibm} and, for odd-mass and odd-odd
nuclei, the interacting boson fermion model
(IBFM)~\cite{iachello1991:ibfm}.  In this context, generalized
seniority states constructed from the $S$ pair and a collective $D$
pair (consisting of like nucleons coupled to angular momentum $2$) are
mapped onto IBM states built from analogous combinations of $s$ and
$d$ bosons~\cite{otsuka1978:ibm2-shell-details,pittel1982:ibm-micro,iachello1987:ibm-shell,otsuka1996:ibm-micro,yoshinaga1996:ibm-micro-sm}.  The need for a fully
microscopic derivation of the IBM Hamiltonian parameters and
transition operators is a longstanding problem.  A well-developed
microscopic mapping would be especially valuable for the
IBFM, where the relevance of
single-particle degrees of freedom is particularly manifest.  For
instance, higher-order mapping from the generalized seniority shell
model space onto the IBFM would facilitate the prediction of
collective effects in $\beta$
decay~\cite{navratil1988:ibfm-beta-a195-a197}. The need is also
emphasized by recent applications of the model to neutrinoless double
beta decay~\cite{barea2009:ibm-doublebeta}, where phenomenological
calibration of the model is not a viable alternative.  For 
microscopic derivation of the IBM and IBFM, matrix elements of various
operators between states of generalized seniority $v\lesssim 6$ are of
interest.

In addition to matrix elements of operators between generalized
seniority states, it is also necessary to compute the overlaps of
these states, since the canonical construction~\cite{gambhir1969:bpm}
leads to a nonorthogonal basis.  The $S$ pair creation operator is
obtained as a linear combination of pair creation operators for each
active single-particle level (\textit{i.e.}, $j$ shell).  The
individual $j$ shells may be treated using the conventional seniority
or quasispin results~\cite{deshalit1963:shell,macfarlane1966:shell}.
Then, the overlaps and matrix elements in the generalized seniority
scheme may in principle be computed from the single-shell results by
combinatorial
arguments~\cite{gambhir1969:bpm,gambhir1971:bpm-ni-sn,pittel1982:ibm-micro}.

However, the combinatorial derivation~\cite{pittel1982:ibm-micro}
rapidly becomes cumbersome with increasing generalized seniority.  It
was noted~\cite{frank1982:ibm-commutator} that commutator methods can
be used to obtain simplifications, but the resulting hybrid approach
could still only be practically applied for
$v\leq2$~\cite{frank1982:ibm-commutator,lipas1990:ibm-micro-escatt}.
The main challenge in proceeding to higher generalized seniority lies
in accounting for the intermediate angular momentum couplings which
occur among the particles not participating in the collective $S$
pair.  Several approaches have been pursued.  The combinatorial approach may
be pushed further by introducing certain simpler intermediate
quantities, essentially norms for uncoupled states, and then
reexpressing the overlaps for the angular-momentum coupled states in
terms of these (however, matrix
elements were not explicitly considered in this formulation)~\cite{vanisacker1986:ibm-cm-micro}.  One may directly carry
out the calculations of overlaps and matrix elements by Wick's theorem
in an uncoupled scheme, subsequently recoupling the
results~\cite{mizusaki1996:ibm-micro}.  An
alternative, indirect construction of the generalized seniority basis
relies upon number projection of generalized-quasiparticle states
involving complex parameters.  Matrix elements are then evaluated as
contour integrals in the complex
plane~\cite{ottaviani1969:np-tda,bonsignori1978:tda-sn,allaart1988:bpm}.
These latter methods have been applied for $v\lesssim 4$.

In the present work, a systematic calculational scheme is established,
in which recurrence relations for the matrix elements and overlaps are
derived by angular-momentum coupled commutator
methods~\cite{french1966:multipole,chen1993:wick-coupled}.  This is
essentially a Wick's theorem approach, which, however, retains the
angular-momentum coupled structure.  General expressions are obtained
by which the one-body operator matrix elements and the overlaps for
states of generalized seniority $v$ and pair number $N$ are expressed
in terms of commutators which yield matrix elements and overlaps
involving lower values of $v$ and/or $N$.  The explicit recurrence
relations to be used in the numerical calculations for a
\textit{specific} value of the generalized seniority are obtained from
these generic expressions by a straightforward procedure involving
repeated application of a commutator product rule.  The matrix
elements of two-body operators can be computed directly in terms of
these basic one-body operator matrix elements and overlaps.  An
essential motivation for the present approach is that calculation can
proceed to higher generalized seniority in a methodical fashion,
through systematic application of this coupled commutator product
rule, and that the process is readily amenable to
automation~\cite{chen1993:wick-clusters}.  For illustration, the
explicit recurrence relations are given for matrix elements and
overlaps involving low generalized seniority states, sufficient for
generalized seniority shell model calculations with $v\leq3$.  In
particular, calculation of the matrix elements of a two-body
Hamiltonian between states of $v=3$ requires one-body operator matrix
elements for $v\leq 4$ and overlaps for $v\leq 5$.

After a review of definitions (Sec.~\ref{sec-defn}), the general
commutation scheme is outlined (Sec.~\ref{sec-recur-scheme}), and the
underlying commutator algebra results are summarized
(Sec.~\ref{sec-comm}).  A coupled set of recurrence relations for the
one-body operator matrix elements (Sec.~\ref{sec-recur-Gamma}) and
overlaps (Sec.~\ref{sec-recur-Phi}) are obtained.  Matrix elements of
two-body operators are then expressed in terms of these quantities
(Sec.~\ref{sec-tbo}).

\section{Definitions}
\label{sec-defn}

First, let us review the definitions necessary for the generalized
seniority scheme.  Within a single major shell, the angular-momentum
label is sufficient to uniquely specify a single-particle level, so
let $\Cd(c,\gamma)$ be the creation operator for a particle in the state
of angular momentum $c$ and $z$-projection quantum number
$\gamma$.\footnote{For simplicity of notation, we do not distinguish
between the level $c$ and its angular momentum $j_c$. Also note that,
following Ref.~\cite{chen1993:wick-coupled}, we denote operators by
capital roman letters, angular momenta by lower case roman letters,
and angular-momentum $z$-projection quantum numbers, when needed, by
the corresponding lower case Greek letter (\textit{e.g},
$A^{a}_\alpha$, $B^{b}_\beta$, $C^{c}_\gamma$).}  The angular-momentum
coupled product of two spherical tensor operators is defined by
$(A^a\times B^b)^c_\gamma=\sum_{\alpha\beta}\tcg a\alpha b\beta
c\gamma A^a_\alpha B^b_\beta$, and we follow the time reversal phase
convention
$\tilde{A}^a_\alpha=(-)^{a-\alpha}A^a_{-\alpha}$~\cite{varshalovich1988:am}.
Then the angular-momentum coupled pair creation operator is defined by
\begin{equation}
\label{eqn-defn-A}
\Ad(abe)=(\Cd(a)\times\Cd(b))^e,
\end{equation}
and its time-reversed adjoint is $\At(abe)\equiv
\widetildedagger{(\Ad(abe))}=-(\Ct(a)\times\Ct(b))^e$.

The state of zero generalized
seniority is defined by the $S$-pair condensate,
$\tket{\S(N)}=\Sd(N)\vket$, where the collective $S$ pair in the generalized seniority scheme is defined
by
\begin{equation}
\label{eqn-defn-S}
S^\dagger=\sum_c \alpha_c  \frac{\hat{c}}{2} \Ad(cc0),
\end{equation}
with $\hat{c}=(2c+1)^{1/2}$.  The amplitudes $\alpha_c$ are
conventionally taken subject to the normalization condition $\sum_c
(2c+1)\alpha_c^2=\sum_c(2c+1)$.  They may be obtained by a variety of
prescriptions~\cite{gambhir1969:bpm,talmi1971:shell-seniority,scholten1983:ibm2-microscopic-majorana,otsuka1993:ibm2-ba-te-microscopic,mizusaki1996:ibm-micro},
for instance, variationally so as to minimize the energy expectation
value $\tme{\S(N)}{H}{\S(N)}$.

More generally, a
state of generalized seniority $v$ is constructed as $\tket{\S(N)
\F(f)}=\Sd(N)\Fd(f)\vket$, where the fermion cluster $\Fd(f)$ is a
product of $v$ creation operators, coupled to total angular
momentum $f$.  A complete set of such clusters is constructed by
successive couplings of the form $\F(f)=((A^\dagger\times
A^\dagger)\cdots\times A^\dagger)^f$ for $v$ even or $\F(f)=(((A^\dagger\times
A^\dagger)\cdots\times A^\dagger)\times C^\dagger)^f$ for $v$ odd.
In particular, the conventional microscopic
interpretation~\cite{iachello1987:ibm-shell} of the IBM is formulated in terms of
generalized seniority states involving the collective $D$ pair, defined as a
general linear combination
\begin{equation}
\label{eqn-defn-D}
D^\dagger=\sum_{\substack{ab\\a\leq
b}}\frac{\beta_{ab}}{\deltasqrt{ab}} \Ad(ab2),
\end{equation}
of pairs of angular momentum $2$.  The states of interest for mapping
from the shell model to the IBM, through the Otsuka-Arima-Iachello
(OAI)
mapping~\cite{otsuka1978:ibm2-shell-details,pittel1982:ibm-micro,iachello1987:ibm-shell,otsuka1996:ibm-micro,yoshinaga1996:ibm-micro-sm},
are of the form $\tket{\S(N)}$, $\tket{\S(N)D}$, $\tket{\S(N)(D D)^f}$,
\textit{etc.}  For the analogous mapping~\cite{yoshinaga2000:gssm-ibfm} to the 
IBFM, the
relevant states are $\tket{\S(N)\C(a)}$, $\tket{\S(N)(D\C(a))^f}$,
\textit{etc.}  Any matrix element or overlap involving $D$ pairs may
be expanded in terms of those for the elementary pairs $\A(ab2)$,
through~(\ref{eqn-defn-D}).

The matrix elements of most immediate interest are those
of one-body and two-body operators.  Any spherical tensor one-body operator
may be expressed in terms of the elementary one-body multipole operators 
\begin{equation}
\label{eqn-defn-T}
\T(rst)=(\Cd(r)\times\Ct(s))^t,
\end{equation}
through the usual second-quantized realization, which in
angular-momentum coupled form becomes
$U^u=-\sum_{ab}\hat{u}^{-1}\trme{a}{U^u}{b}\,\T(abu)$.\footnote{However,
in comparing with, \textit{e.g.},
Ref.~\cite{suhonen2007:nucleons-nucleus}, note that the overall sign
of this expression varies depending upon the time-reversal phase
convention, presently
$\tilde{A}^a_\alpha=(-)^{a-\alpha}A^a_{-\alpha}$.} Thus, we must
consider matrix elements of $\T(rst)$ taken between states of the
generalized seniority scheme, which are of the form
$\trme{\S(N)\G(g)}{\T(rst)}{\S(N)\F(f)}$, where $\F(f)$ and $\G(g)$
represent clusters of $v$ nucleons not participating in the collective
$S$ pairs.  Since the generalized seniority states are not
orthonormal, as canonically constructed in terms of clusters
above~\cite{otsuka1996:ibm-micro}, it is also necessary to compute the
norms and overlaps $\toverlap{\S(N)\G(f)}{\S(N)\F(f)}$.

The creation operators for the generalized seniority basis states, as
defined above, have a uniform structure consisting of sequentially
coupled pair creation operators (with a single additional creation
operator in the case of odd $v$).  Therefore, it is more compact and
readable to simply label these states by the single-particle level
labels and the angular momenta for intermediate couplings, that is, as
$\tket{\S(N)} \equiv \Sd(N)\vket$, $\tket{\S(N) \pclusterC(c)}
\equiv\Sd(N) \dclusterC(c)\vket$,
$\tket{\S(N)\pclusterA(abe)} \equiv \Sd(N)\dclusterA(abe)\vket$, 
$\tket{\S(N) \pclusterAC(abeig)}\equiv \Sd(N) \dclusterAC(abeig)\vket$,
$\tket{\S(N) \pclusterAA(abeijmg)} \equiv \Sd(N)
\dclusterAA(abeijmg)\vket$, \textit{etc.}  

Only matrix elements and overlaps involving bra and ket states of
\textit{equal} generalized seniority need be calculated explicitly
through recurrence relations.  To facilitate writing (and discussing)
the recurrence relations, we furthermore introduce the symbol
$\OLAP_N^{(v)}[\cdots]$ for the overlaps of states of equal
generalized seniority $v$, as
\begin{equation}
\label{eqn-defn-Phi}
\begin{aligned}
  \OLAPzero(N) &\equiv \toverlap{\S(N)}{\S(N)}
\\
  \OLAPone(N)[c|c] &\equiv \toverlap{\S(N) \pclusterC(c)}{\S(N)
  \pclusterC(c)}
\\
  \OLAPtwo(N)[cde|abe] &\equiv \toverlap{\S(N) \pclusterA(cde)}{\S(N)
  \pclusterA(abe)}
\\
  \OLAPthree(N)[cdfjg|abeig] &\equiv \toverlap{\S(N)
  \pclusterAC(cdfjg)}{\S(N) \pclusterAC(abeig)}
\\
  \OLAPfour(N)[cdfklng|abeijmg] &\equiv \toverlap{\S(N)
  \pclusterAA(cdfklng)}{\S(N) \pclusterAA(abeijmg)},
\end{aligned}
\end{equation}
\textit{etc.}  Similarly, we introduce $\OBME_N^{(v)}[\cdots]$ for the
one-body operator matrix elements between these states, \textit{e.g.},
\begin{equation}
\label{eqn-defn-Gamma}
\begin{aligned}
  \OBMEfour(N)[cdfklnh|rst|abeijmg] &\equiv \trme{\S(N)
  \pclusterAA(cdfklnh)}{\T(rst)}{\S(N) \pclusterAA(abeijmg)}.
\end{aligned}
\end{equation}
These definitions are intended to follow the bracket notation as
closely as possible, while clearly exhibiting the labels $v$ and $N$
with respect to which the recurrence will be carried out, and also
more simply laying out the arguments of $\OLAPsymbol{v}{N}$ and $\OBMEsymbol{v}{N}$
considered as symbols in the recursive computational scheme.

Matrix elements or overlaps involving states of \textit{unequal}
generalized seniority can readily be expressed in terms of these by
expanding one or more $S$ pairs in terms of the pairs
$\lclusterA(aa0)$,
\textit{i.e.}, created by $\dclusterA(aa0)$, for individual $j$
shells, according to the definition~(\ref{eqn-defn-S}),
\textit{e.g.},
\begin{equation}
\label{eqn-v-upgrade-example}
\toverlap{\S(N)\pclusterA(cdg) }{\S(N-1)
\pclusterAA(abeijmg)}=\sum_k \alpha_k \frac{\hat{k}}{2}\OLAPfour(N-1)[kk0cdgg|abeijmg].
\end{equation}
Here we notice that, although the pair $\lclusterA(kk0)$ carries
\textit{seniority} zero, it is not the \textit{collective} $S$ pair
and therefore carries a \textit{generalized} seniority of $2$.

It should be noted that the $\OBME^{(v)}_N$ and $\OLAP^{(v)}_N$
obey a variety of symmetry relations under rearrangement of the
arguments, and therefore they need not all be calculated
independently.  For instance, different orderings of single-particle
labels within the bra or ket are related by,
\textit{e.g.}, 
$\Ad(bae)=-\theta(abe)\Ad(abe)$ and
$\dclusterAA(abeijmg)=\theta(emg)\dclusterAA(ijmabeg)$, where
\begin{equation}
\label{eqn-defn-theta}
\theta(abc\cdots)=(-)^{a+b+c+\cdots}.
\end{equation}
Thus,
for instance,
\begin{equation}
\label{eqn-symm-Phi4-ket-pair}
\OBMEfour(N)[cdfklnh|rst|abeijmg]
=-\theta(abe)\OBMEsymbol{4}{N}[(cd)^f(kl)^{n\,h}|\lclusterA(rst)|(\underarrow{leftright}{ba})^e(ij)^{m\,g}],
\end{equation}
interchanging particles within a pair, as indicated by the arrows, or
\begin{equation}
\label{eqn-symm-Phi4-ket-cluster}
\OBMEfour(N)[cdfklnh|rst|abeijmg]
=\theta(emg)\OBMEsymbol{4}{N}[(cd)^f(kl)^{n\,h}|\lclusterA(rst)|(i\underarrow{leftright}{j)^m(a}b)^{e\,g}],
\end{equation}
interchanging pairs within a cluster.
Moreover, the
property of the reduced matrix element under complex conjugation,
$\trme{c}{B^b}{a}^*=(-)^{a+b-c}\trme{a}{\tilde{B}^{b\,\dagger}}{c}$,
gives the relation
\begin{equation}
\label{eqn-Gamma-symm-adjoint}
\trme{\S(N)\G(g)}{\T(rst)}{\S(N)\F(f)}=-(-)^{r+s+g-f}
\trme{\S(N)\F(f)}{\T(srt)}{\S(N)\G(g)},
\end{equation}
where we note that
$\tilde{\tilde{C}}^\dagger_c=-\Cd(c)$, and consequently
$\tilde{T}_{rs}^{t\,\dagger}=-\theta(rst)\T(srt)$.
Thus, for instance,
\begin{equation}
\label{eqn-symm-Phi4-adjoint}
\OBMEfour(N)[cdfklnh|rst|abeijmg]
=-\theta(rsgh)\OBMEsymbol{4}{N}[(ab)^e\underarrow{leftright}{(ij)^{m\,g}|(sr)^t|(cd)^f}(kl)^{n\,h}],
\end{equation}
interchanging bra and ket and conjugating the operator.


\section{General scheme for recurrence}
\label{sec-recur-scheme}

Recurrence relations for the reduced matrix elements and overlaps can
be naturally derived by commutator methods.  These matrix elements and
overlaps must first be expressed as vacuum expectation values of
coupled products of creation and annihilation operators.  The
operators are then reordered, by making use of their commutation (or
anticommutation) relations, until all the resulting terms themselves
represent matrix elements or overlaps of generalized seniority states.
Since the commutators (or anticommutators) yield terms with
\textit{fewer} total creation and annihilation operators, the matrix
elements and overlaps resulting from these terms will
automatically involve states of lower pair number or generalized seniority, thus giving rise to 
relations which are recursive with respect to
$N$ or $v$.

In general, consider two states
$\tket{A^a_\alpha}=A^{a\,\dagger}_\alpha\vket$ and
$\tket{C^c_\gamma}=C^{c\,\dagger}_\gamma\vket$, obtained
by the action of some operators $A^{a\,\dagger}_\alpha$ and $C^{c\,\dagger}_\gamma$
on the vacuum, for instance, coupled products of fermion creation
operators.  Then the reduced matrix element of an operator $B^b$ can
be reexpressed as the
vacuum expectation value
\begin{equation}
\label{eqn-rme-vev}
\trme{C^c}{B^b}{A^a}=(-)^{a-b-c}\vev{[\tilde{C}^c\times(B^b\times A^{a\,\dagger})^c]^0_0}
\end{equation}
of a scalar triple product of
operators.\footnote{Relation~(\ref{eqn-rme-vev}) is readily derived
from the identity
\smash{$\protect\protect\trme{c}{B^b}{a}=\sum_{\alpha\beta\gamma}(-)^{c-\gamma}\protect\smallthreej{c}{b}{a}{-\gamma}{\beta}{\alpha}\protect\tme{c\gamma}{B^b_\beta}{a\alpha}$}~\cite{edmonds1960:am}
and is the operator analog of, \textit{e.g.}, (14.16) of Ref.~\cite{deshalit1963:shell}.
We follow the normalization
and phase convention of Refs.~\cite{edmonds1960:am,varshalovich1988:am} for the Wigner-Eckart theorem, \textit{i.e.}, 
\smash{$\protect\tme{c\gamma}{B^b_\beta}{a\alpha}=(-)^{2b}\hat{c}^{-1}
\protect\tcg{a}{\alpha}{b}{\beta}{c}{\gamma}
\protect\trme{c}{B^b}{a}$}.
} 
The overlap $\toverlap{B^a}{A^a}\equiv\toverlap{B^a_\alpha}{A^a_\alpha}$ of two
states can similarly be evaluated as the vacuum expectation value
\begin{equation}
\label{eqn-overlap-vev}
\toverlap{B^a}{A^a}=\hat{a}^{-1}\vev{(\tilde{B}^a\times A^{a\,\dagger})^0_0}.
\end{equation}

The matrix element of the one-body operator $\T(rst)$ between states
of generalized seniority $v$ is therefore given by an expression of the form
\begin{equation}
\label{eqn-rme-Gamma}
\trme{\S(N)\G(g)}{\T(rst)}{\S(N)\F(f)}
=(-)^{f-t-g}
\vev{(\Gt(g)\St(N)\times\T(rst)\times\Sd(N)\Fd(f))^0},
\end{equation}
where $\Fd(f)$ and $\Gd(g)$ represent two clusters consisting of $v$
fermionic creation operators.  
The overlap of two states of generalized seniority $v$ is similarly
\begin{equation}
\label{eqn-overlap-Phi}
\toverlap{\S(N)\G(f)}{\S(N)\F(f)}
=\hat{f}^{-1}
\vev{(\Gt(f)\St(N)\times\Sd(N)\Fd(f))^0},
\end{equation}
where again $\Fd(f)$ and $\Gd(f)$ represent two clusters consisting of
$v$ fermionic creation operators, now with the same angular momentum
$f=g$. 

To outline the general scheme for deriving the recurrence relation
for a matrix element $\OBME^{(v)}_N$,
let us momentarily suppress the details of angular-momentum coupling,
the single-particle level labels, and the various numerical
coefficients.  To highlight the rearrangement taking place in each
step, the factors to be reordered are indicated by an underbrace, and
the general form of the commutator introduced by this reordering (if
the factors do not freely commute) is shown schematically underneath.  It suffices to
note that commutation of an operator $T$ or $\tilde{A}$ through $\Sd(N)$ will yield terms
of the form $[T,\Sd(N)]\sim A^\dagger \Sd(N-1)$
[see~(\ref{eqn-comm-SdNT}) in Sec.~\ref{sec-recur-Gamma}] or $[\tilde{A},\Sd(N)]\sim
\Sd(N-1) + T \Sd(N-1) + \Sd(N-1)T$ [see~(\ref{eqn-comm-SdNAt}) in Sec.~\ref{sec-recur-Gamma}].  The time reversed adjoint expressions
(see Sec.~\ref{sec-comm}) apply for commutation through $\St(N)$.  Furthermore, new clusters
will arise from commutation through the clusters $F$ and $G$,
specifically, with creation operators defined by
$E^\dagger\sim[T,G^\dagger]$, $H^\dagger\sim[T,F^\dagger]$, and
$I^\dagger\sim[\tilde{A},G^\dagger]$, to be calculated as described in
Sec.~\ref{sec-comm}.

Beginning with the reduced matrix element
$\OBME^{(v)}_N=\trme{\S(N) G}{T}{\S(N) F}$, expressed as a vacuum expectation
value as in~(\ref{eqn-rme-Gamma}), we set out to commute the one-body
operator to the right, where it will annihilate the vacuum.  Thus,
factors must be reordered as
\begin{equation}
\label{eqn-recur-Gamma-schematic-1}
\begin{aligned}
\OBME^{(v)}_N&\sim\vev{(\tilde{G}\St(N)) \,
\underbracevalue{T \, (\Sd(N)}{A^\dagger\Sd(N-1)} 
F^\dagger) }
\\ 
&\sim
\vev{(\tilde{G}\St(N))\,(\Sd(N)
\underbracevalue{T F^\dagger}{\equiv H^\dagger})
} +\vev{
\tilde{G}\St(N)A^\dagger\Sd(N-1)F^\dagger
}
\\
&\sim
\vev{(\tilde{G}\St(N))\,(\Sd(N)
H^\dagger)
} +\vev{
\tilde{G}\St(N)A^\dagger\Sd(N-1)F^\dagger
}.
\end{aligned}
\end{equation}
The first term is recognized as the overlap $\toverlap{\S(N) G}{\S(N)
H}\sim\vev{ (\tilde{G}\St(N))(\Sd(N)H^\dagger)}$.  Note that the
cluster $H$ again contains $v$ fermions.  To evaluate
the second term, we must commute $A^\dagger$ to the left, where it
will eventually annihilate the vacuum, yielding
\begin{multline}
\label{eqn-recur-Gamma-schematic-2}
\vev{
\tilde{G}
\underbracevalue{
\St(N)A^\dagger}{\St(N-1)+T\St(N-1)+\St(N-1)T}
\Sd(N-1)F^\dagger
}
\\
\begin{lgathered}[t]
\sim
\vev{(
\underbracevalue{\tilde{G}
A^\dagger}{\equiv\tilde{I}}\St(N))\,(\Sd(N-1)F^\dagger)}
+\vev{(\tilde{G}\St(N-1))\,(\Sd(N-1)F^\dagger)}
\\
\quad\qquad
+\vev{(\underbracevalue{\tilde{G}T}{\equiv\tilde{E}}\St(N-1))\,(\Sd(N-1)F^\dagger)}
+\vev{(\tilde{G}\St(N-1))\,T\,(\Sd(N-1)F^\dagger)}
\\
\sim
\vev{(\tilde{I}\St(N))\,(\Sd(N-1)F^\dagger)}
+\vev{(\tilde{G}\St(N-1))\,(\Sd(N-1)F^\dagger)}
\\
\quad\qquad
+\vev{(\tilde{E}\St(N-1))\,(\Sd(N-1)F^\dagger)}
+\vev{(\tilde{G}\St(N-1))\,T\,(\Sd(N-1)F^\dagger)}.
\end{lgathered}
\end{multline}
The first term is recognized as the overlap $\toverlap{\S(N-1)
(SI)}{\S(N-1) F}$, the
second term as $\toverlap{\S(N-1) G}{\S(N-1) F}$, the third term as
$\toverlap{\S(N-1) E}{\S(N-1) F}$, 
and the final term as $\trme{\S(N-1) G}{T}{\S(N-1) F}$.  Note that each of the
clusters $E$, $H$, and $SI$ [where the $S$ pair is expanded as in~(\ref{eqn-v-upgrade-example})] again contains $v$ fermions.
Therefore, the resulting
recurrence relation has the form
\begin{equation}
\label{eqn-recur-Gamma-symb}
\OBME^{(v)}_N\sim \OLAP^{(v)}_N+\OBME^{(v)}_{N-1}+\OLAP^{(v)}_{N-1}.
\end{equation}

Similarly, a reduction in seniority $v$ is then provided through the recurrence
relation for the overlap $\OLAP^{(v)}_N=\toverlap{\S(N) G}{\S(N) F}$.
The overlap is expressed as a vacuum expectation value as in~(\ref{eqn-overlap-Phi}), and a fermion
is ``decoupled'' from each of the initial clusters $F$ and
$G$, leaving behind a subcluster of
reduced seniority, denoted by $H$ or $I$, \textit{i.e.}, $F^\dagger\sim C^\dagger H^\dagger$
and $G^\dagger\sim C^\dagger I^\dagger$.  These fermion operators are
then migrated inwards and recoupled to constitute a one-body operator,
thereby yielding the reduced matrix element of a one-body
operator, but now of lower
seniority.  Schematically,
\begin{equation}
\label{eqn-recur-Phi-schematic}
\begin{aligned}
\OLAP^{(v)}_N&\sim\vev{(\tilde{G}\St(N))\,(\Sd(N) F^\dagger)}\\
&\sim\vev{(\tilde{I}\underbracenovalue{\tilde{C}\St(N)})\,(\underbracenovalue{\Sd(N)
C^\dagger} H^\dagger)}\\
&\sim\vev{(\tilde{I}\St(N))\,(\underbracevalue{\tilde{C}C^\dagger}{1})\,(\Sd(N)
H^\dagger)}\\ &\sim\vev{(\tilde{I}\St(N))\,T\,(\Sd(N) H^\dagger)} +
\vev{(\tilde{I}\St(N))\,(\Sd(N) H^\dagger)},
\end{aligned}
\end{equation}
where the latter term arises from the
canonical anticommutator of $\tilde{C}$ and $C^\dagger$.  The 
result is recognized as consisting of a matrix element and an overlap, respectively,
of seniority $v-1$.  The recurrence relation thus has the form
\begin{equation}
\label{eqn-recur-Phi-symb}
\OLAP^{(v)}_N\sim \OBME^{(v-1)}_{N}+\OLAP^{(v-1)}_{N}.
\end{equation}
\begin{figure}[t]
\begin{center}
\includegraphics*[width=0.6\hsize]{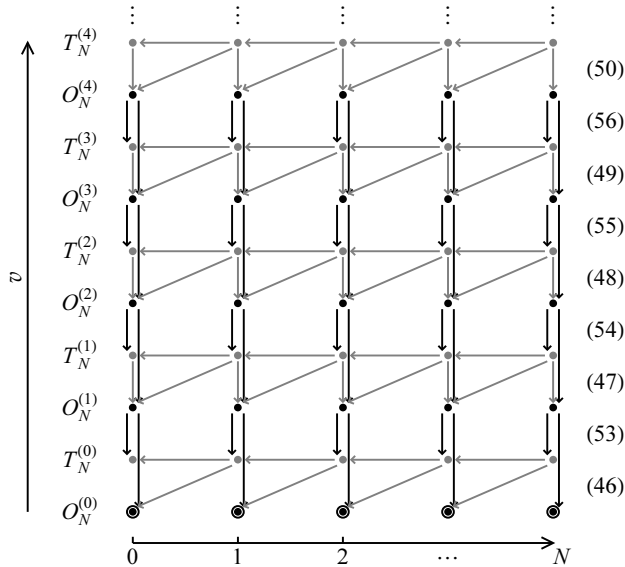}
\end{center}
\caption{Schematic recurrence network for the calculation of one-body
operator reduced matrix elements $\OBME^{(v)}_N$ and overlaps
$\OLAP^{(v)}_N$ in the generalized seniority scheme, as described
by~(\ref{eqn-recur-Gamma-symb}) and~(\ref{eqn-recur-Phi-symb}).
Equation numbers are indicated at right for the explicit recurrence
relations (Secs.~\ref{sec-recur-Gamma} and~\ref{sec-recur-Phi})
appropriate to each generalized seniority.  The seed values
$\OLAPzero(N)$ (circled) are given by~(\ref{eqn-Phi0-comb}).  Each point (except
for the seed values) represents multiple quantities
$\OBME^{(v)}_N[\cdots]$ or $\OLAP^{(v)}_N[\cdots]$, distinguished by
the single-particle level and
angular-momentum coupling labels.  }
\label{fig-recur}
\end{figure}

The recurrence relations of the form~(\ref{eqn-recur-Gamma-symb})
and~(\ref{eqn-recur-Phi-symb}), to be set out in detail in Secs.~\ref{sec-recur-Gamma}
and~\ref{sec-recur-Phi}, allow any matrix element $\OBME^{(v)}_N$ or
overlap $\OLAP^{(v)}_N$ to ultimately be reexpressed in terms 
of the norms of states $\tket{\S(k)}$ of generalized seniority zero,
\textit{i.e.}, the
$\OLAPzero(k)$ ($k=0,1,\ldots,N$).  These constitute the seed values
for the recurrence.  The recurrence network for evaluating a given $\OBME^{(v)}_N$ or
$\OLAP^{(v)}_N$ is shown schematically in
Fig.~\ref{fig-recur}.  The actual recurrence relations involve sums
over single-particle indices and over angular momenta for intermediate
couplings.

The overlaps $\OLAPzero(N)$ are
given~\cite{gambhir1969:bpm,pittel1982:ibm-micro}, in terms of the
coefficients $\alpha_c$ defining the $S$ pair in~(\ref{eqn-defn-S}),  by
\begin{equation}
\label{eqn-Phi0-comb}
\OLAPzero(N)=(N!)^2\sumpart{M}{N,D} 
\biggl[\prodsp{c} \alpha_c^{2M_c}\binom{\Omega_c}{M_c}\biggr].
\end{equation}
The sum is over all ordered partitions $M=(M_1,M_2,\ldots,M_D)$ of the integer $N$ into $D$ terms $M_c$ ($\sum_c M_c=N$), where $D$
is the number of active single-particle levels.  That is, the sum is
over all possible ways in which $N$ pairs may be distributed over the
$D$ levels.  Only Pauli-allowed
occupancies, 
$M_c\leq\Omega_c$, contribute, where $\Omega_c=(2c+1)/2$ is the pair
degeneracy of the level.

\section{Commutator algebra}
\label{sec-comm}

In order to carry out the commutation scheme for the elementary
one-body operator matrix elements $\OBME^{(v)}_N$, as outlined in
Sec.~\ref{sec-recur-scheme}, it is necessary to reorder various
factors within spherical tensor coupled products.  For each value of
the generalized seniority $v$ being considered, it is seen
from~(\ref{eqn-recur-Gamma-schematic-1})--(\ref{eqn-recur-Gamma-schematic-2})
that commutators of the form $[\tilde{A},\Fd(f)]$ and $[T,\Fd(f)]$
will be needed, where $\Fd(f)$ is the fermionic cluster creation
operator containing $v$ creation operators.  That is, $F^\dagger=(A^\dagger\times\cdots)\times A^\dagger$ or $((A^\dagger\times\cdots)\times A^\dagger)\times
C^\dagger$.  A straightforward
approach is provided by the coupled commutator methods of Chen
\textit{et al.}~\cite{chen1993:wick-coupled}, as summarized in this
section.  All such commutators for
$v\leq4$ (some of which may be found in
Refs.~\cite{french1966:multipole,gambhir1969:bpm,frank1982:ibm-commutator,chen1993:wick-coupled,allaart1988:bpm})
are collected in coupled form below.

The spherical tensor \textit{coupled
commutator}~\cite{french1966:multipole} is defined by
\begin{equation}
\label{eqn-comm-defn}
[A^a,B^b]^e_\varepsilon=\sum_{\alpha\beta}\tcg a\alpha b\beta
e\varepsilon [A^a_\alpha,B^b_\beta].
\end{equation}
Since fermionic creation operators obey a canonical
\textit{anticommutation} relation, it is more useful in the present
context to introduce the
\textit{graded coupled commutator}, developed by Chen \textit{et al.}~\cite{chen1993:wick-coupled}, in which the commutation 
bracket is defined by
\begin{equation}
\label{eqn-defn-graded}
[A^a_\alpha,B^b_\beta]=A^a_\alpha B^b_\beta-\theta_{ab} B^b_\beta
A^a_\alpha,
\end{equation}
that is, the
\textit{commutator} ($\theta_{ab}=+1$) if either angular momentum $a$
or $b$ is integer (quasi-bosonic) but the
\textit{anticommutator} ($\theta_{ab}=-1$) if both angular momenta $a$
and $b$ are odd half-integer (quasi-fermionic).  [Note
$\theta_{ab}\not\equiv\theta(ab)$.] The quantity
$[A^a,B^b]^e$ itself constitutes a spherical tensor, indeed, manifestly so,
since it may be expressed
in terms of coupled products of $A^a$ and $B^b$ as
\begin{equation}
\label{eqn-relation-coupled}
[A^a,B^b]^e=(A^a\times B^b)^e-\theta_{ab}(-)^{e-a-b}(B^b\times A^a)^e.
\end{equation}
This provides the basic relation for reordering factors within coupled products.
The use of coupled commutators thus circumvents the cumbersome
process of uncoupling products of operators, commuting the components,
and recoupling the result.  The commutator has the symmetry (or antisymmetry) property
\begin{equation}
\label{eqn-symm-coupled}
[B^b,A^a]^e=-\theta_{ab}(-)^{e-a-b}[A^a,B^b]^e
\end{equation}
under interchange of its arguments.

The canonical anticommutation property for fermionic creation and
annihilation operators,
$[\C(a,\alpha),\Cd(b,\beta)]=\delta_{ab}\delta_{\alpha\beta}$ (recall
this is the graded commutator), is written in coupled form as
\begin{equation}
\label{eqn-comm-canonical}
[\Ct(a),\Cd(b)]^e=\hat{a}\delta_{ab}\delta_{e0}.
\end{equation}
Any coupled commutator of more complicated operators can be reduced
to~(\ref{eqn-comm-canonical}) by repeated application of the coupled
commutator product rule
\begin{multline}
\label{eqn-comm-prod}
[A^a,(C^c\times D^d)^f]^g=\sum_h \usixj cdfgah ([A^a,C^c]^h\times
D^d)^g
\\
+\sum_h(-)^{d+g-f-h}\theta_{ac} \usixj cdfagh
(C^c\times[A^a,D^d]^h)^g,
\end{multline}
which applies for arbitrary spherical tensor operators
$A^a$, $C^c$, and $D^d$.
The quantity in brackets is the unitary $6$-$j$ symbol,
\begin{equation}
\label{eqn-defn-usixj}
\usixj abcdef = (-)^{a+b+d+e} {\hat c \hat f} \sixj abcdef,
\end{equation}
commonly denoted by $U(abed;cf)$.

Commutators of the one-body operator $T$ with the creation
operators for fermion clusters are obtained by repeated
application of the product rule~(\ref{eqn-comm-prod}):
\begin{align}
\label{eqn-comm-T-C}
[\T(abe),\dclusterC(c)]^d&=\theta(cde)\frac{\hat e}{\hat d}
\delta_{bc}\delta_{ad}\Cd(d)
\\
\label{eqn-comm-T-A}
[\T(abe),\dclusterA(cdf)]^g&=\P(cdf)\theta(cdeg)\frac{\hat e \hat
f}{\hat d \hat g} \delta_{bd}\usixj fegacd \Ad(acg)
\\
\label{eqn-comm-T-AC}
[\T(rst),\dclusterAC(abecg)]^h&=
\begin{lgathered}[t]
\sum_x\P(abe)
\theta(abtx)\frac{\hat
e \hat t}{\hat b \hat x} \delta_{bs}
\\
\qquad\times 
\usixj ecghtx \usixj etxrab \dclusterAC(raxch)
\\
+\theta(ght)\frac{\hat t}{\hat r} \delta_{cs}\usixj ecgthr
\dclusterAC(aberh)
\end{lgathered}
\\
\label{eqn-comm-T-AA}
[\T(rst),\dclusterAA(abecdfg)]^h&=
\begin{lgathered}[t]
\sum_x[1+\theta(efg)\PP{abe}{cdf}]\P(cdf)
\\
\qquad\times
\theta(cdfght)\frac{\hat f
\hat t}{\hat d \hat x} \delta_{ds}
\usixj efgthx \usixj ftxrcd \dclusterAA(abercxh).
\end{lgathered}
\end{align}
The exchange symbol $\P(abe)$
(\textit{e.g.}, Refs.~\cite{french1966:multipole,gambhir1969:bpm})
indicates that the following expression should be expanded to consist
of
\textit{two} terms related by interchange of the indices $a$ and $b$,
as
\begin{equation}
\label{eqn-def-P}
\P(abe)\,f(a,b)=f(a,b)-\theta(abe)f(b,a),
\end{equation}
and 
$\PP{abc}{xyz}$ indicates interchange of the indices
$abc$ with $xyz$.

Commutators of the pair annihilation operator $\tilde{A}$ with the creation
operators for fermion clusters are likewise obtained by 
application of the product rule~(\ref{eqn-comm-prod}):
\begin{align}
\label{eqn-comm-At-C}
[\At(abe),\dclusterC(c)]^d&=\frac{\hat e}{\hat d}\delta(ab;cd;e)\Ct(d)\\
\label{eqn-comm-At-A}
[\At(abe),\dclusterA(cdf)]^g&=\begin{multlined}[t]
\hat{e}\delta_{ef}\delta_{g0}\delta(ab;cd;e)
\\
-\P(cdf)\P(abe)\theta(adef)\frac{\hat e \hat f}{\hat b \hat g}
\delta_{bc}
\usixj efgdab \T(dag)
\end{multlined}
\\
\label{eqn-comm-At-AC}
[\At(rst),\dclusterAC(abecg)]^h&=\begin{multlined}[t]
 -\theta(ght)\frac{\hat g}{\hat
h}\delta_{et}\delta_{ch}\delta(ab;rs;e)\Cd(h)
\\
+\P(rst)\P(abe)\theta(egrt)\frac{\hat e \hat t}{\hat a \hat
b}\delta_{ah}\delta_{br}\delta_{cs}
\usixj cegatb \Cd(h) + \calNt
\end{multlined}\\
\label{eqn-comm-At-AA}
[\At(rst),\dclusterAA(abecdfg)]^h&=\begin{multlined}[t]
[1+\theta(efg)\PP{abe}{cdf}]\frac{\hat g}{\hat
h}\delta_{eh}\delta_{ft}\delta(cd;rs;f)\Ad(abh)
\\
-\P(rst)\P(cdf)\P(abe){\hat e \hat f \hat g \hat
t}\delta_{br}\delta_{ds}
\ninej thgdcfbae \Ad(ach) + \calNt,
\end{multlined}
\end{align}
where we adopt the shorthand notation 
\begin{equation}
\label{eqn-def-delta}
\delta(ab;cd;e)=\delta_{ac}\delta_{bd}-\theta(abe)\delta_{ad}\delta_{bc}.
\end{equation}
The $9$-$j$ symbol arises as a sum over products of $6$-$j$ symbols
from~(\ref{eqn-comm-prod}), by identity~(9.8.5) of
Ref.~\cite{varshalovich1988:am}.  In~(\ref{eqn-comm-At-AC})
and~(\ref{eqn-comm-At-AA}), $\calNt$ represents additional normal-ordered
terms containing at least one annihilation operator, which are not explicitly needed
for the present calculations, since in Sec.~\ref{sec-recur-Gamma}
these commutators act
directly on the vacuum.

Commutators involving the time-reversed adjoint of the creation
operator for a fermionic cluster also arise
in~(\ref{eqn-recur-Gamma-schematic-2}), from $\Gt(g)$.  However, these
may be converted into commutators of the type considered above by use of the relation
\begin{equation}
\label{eqn-prod-dt}
\widetildedagger{{(A^a\times B^b)}^e}=(-)^{e-a-b}(\tilde{B}^{b\,\dagger}\times\tilde{A}^{a\,\dagger})^e,
\end{equation}
and therefore
\begin{equation}
\label{eqn-comm-dt}
\widetildedagger{{[A^a,B^b]}^e}=(-)^{e-a-b}[\tilde{B}^{b\,\dagger},\tilde{A}^{a\,\dagger}]^e.
\end{equation}

\section{Recurrence relations for matrix elements of one-body operators}
\label{sec-recur-Gamma}

The recurrence relations for $\OBME^{(v)}_N$ are obtained by applying
the commutation scheme outlined in Sec.~\ref{sec-recur-scheme} to the
one-body operator reduced matrix element defined
by~(\ref{eqn-rme-Gamma}).  
The requisite commutators are~\cite{frank1982:ibm-commutator} 
\begin{align}
\label{eqn-comm-SdNT}
[\Sd(N),\T(cdf)]^f&=N\alpha_d\Ad(cdf)\Sd(N-1)
\\
\label{eqn-comm-SdNAt}
[\Sd(N),\At(cdf)]^f&=-\alpha_c\hat{c}\delta_{cd}\delta_{f0}N\Sd(N-1)
-\alpha_c N \Sd(N-1) \T(cdf) + \theta(cdf)\alpha_d N\T(dcf) \Sd(N-1),
\end{align}
which follow from~(\ref{eqn-comm-T-A}) and~(\ref{eqn-comm-At-A}),
respectively, by induction on $N$.
The general recurrence relation
\begin{multline}
\label{eqn-recur-Gamma}
\underbracketnote{
\trme{\S(N)\G(g)}{\T(rst)}{\S(N)\F(f)}
} {$\OBME_N^{(v)}[\cdots]$}
=
(-)^{f-t-g}\hat{g}
\underbracketnote{
\toverlap{\S(N)\G(g)}{\S(N)\H(g)}
} {$\OLAP_N^{(v)}[\cdots]$}
-N^2\alpha_s^2
\underbracketnote{
\trme{\S(N-1)\G(g)}{\T(rst)}{\S(N-1)\F(f)}
} {$\OBME_{N-1}^{(v)}[\cdots]$}
\\
-\sum_x\tfrac12 N\alpha_s\alpha_x\hat{f}\hat{x}
\underbracketnote{
\toverlap{\S(N-1)(\A(xx0)\I(f))}{\S(N-1)\F(f)}
} {$\OLAP_{N-1}^{(v)}[\cdots]$}
-N^2\alpha_r\alpha_s\hat{f}
\underbracketnote{
\toverlap{\S(N-1)\E(f)}{\S(N-1)\F(f)}
} {$\OLAP_{N-1}^{(v)}[\cdots]$}
\\-N^2\alpha_r^2\hat{r}\hat{f}\delta_{rs}\delta_{fg}\delta_{t0}
\underbracketnote{
\toverlap{\S(N-1)\G(g)}{\S(N-1)\F(f)}
} {$\OLAP_{N-1}^{(v)}[\cdots]$}
\end{multline}
is obtained, in terms of new fermion clusters defined by the coupled commutators
\begin{equation}
\label{eqn-subclusters-Gamma}
\begin{aligned}
\Ed(f)&=[\T(rst),\Gd(g)]^{f}\\
\Hd(g)&=[\T(rst),\Fd(f)]^{g}\\
\Id(f)&=[\At(rst),\Gd(g)]^{f}.
\end{aligned}
\end{equation}

Note that each state appearing in the matrix element and overlaps on
the right hand side of~(\ref{eqn-recur-Gamma}) is, like the original
states $\tket{\S(N)\F(f)}$ and $\tket{\S(N)\G(g)}$, also of generalized
seniority $v$.  In particular, the cluster creation operators $\Ed(f)$
and $\Hd(g)$ each again contain $v$ fermion creation operators, and,
although $\Id(f)$ contains $v-2$ fermion creation operators, it
appears multiplied by a pair creation operator $\Ad(xx0)$, and the
combination $\Ad(xx0)\Id(f)$ again carries generalized seniority $v$.
The \textit{matrix element} $\OBME_{N}^{(v)}$ has thus been recast in
terms of an
\textit{overlap} $\OLAP_{N}^{(v)}$ of the same $N$ and $v$, as well as  a
matrix element $\OBME_{N-1}^{(v)}$ and overlaps $\OLAP_{N-1}^{(v)}$ of
lower pair number, providing the basis for recursive calculation with
respect to pair number.

For $v=0$: As a simple special case, the general recurrence formula~(\ref{eqn-recur-Gamma})
gives
\begin{equation}
\label{eqn-recur-Gamma0}
\OBMEzero(N)[aa0]=
-N^2\alpha_a^2\OBMEzero(N-1)[aa0] -N^2\alpha_a^2\hat{a}\OLAPzero(N-1).
\end{equation}
This may readily be expressed in closed form, in terms of the
seed values $\OLAPzero(N)$ given 
in~(\ref{eqn-Phi0-comb}), as $\OBMEzero(N)[aa0]=
\sum_{k=0}^{N-1}(-)^{N+k}(N!^2/k!^2)\alpha_a^{2(N-k)}\hat{a}\OLAPzero(k)$.

For $v=1$: The recurrence relation obtained
from~(\ref{eqn-recur-Gamma}), with the identifications
$\Fd(f)\rightarrow\dclusterC(c)$ and $\Gd(g)\rightarrow\dclusterC(d)$,
is
\begin{multline}
\label{eqn-recur-Gamma1}
\OBMEone(N)[d|rst|c]=
-\hat{t}\delta_{cs}\delta_{dr}\OLAPone(N)[d|d]
-N^2\alpha_s^2\OBMEone(N-1)[d|rst|c]
\\
-N^2\theta(rst)\alpha_r\alpha_s\hat{t}\delta_{ds}\delta_{cr}\OLAPone(N-1)[c|c]
-N^2\alpha_r^2\hat{r}\hat{c}\delta_{rs}\delta_{cd}\delta_{t0}\OLAPone(N-1)[c|c].
\end{multline}
The coupled commutators needed  for
evaluation of $\Id(f)$, $\Ed(f)$, and $\Hd(g)$ are given
by~(\ref{eqn-comm-T-C}) and~(\ref{eqn-comm-At-C}).

For $v=2$: The recurrence relation, obtained with
$\Fd(f)\rightarrow\dclusterA(abe)$ and
$\Gd(g)\rightarrow\dclusterA(cdf)$, is
\begin{multline}
\label{eqn-recur-Gamma2}
\OBMEtwo(N)[cdf|rst|abe]
\\
=
\begin{lgathered}[t]
\P(abe)\theta(abe)\frac{\hat{e}\hat{t}}{\hat{b}}\delta_{bs}
\usixj etfrab \OLAPtwo(N)[cdf|raf]
-N^2\alpha_s^2\OBMEtwo(N-1)[cdf|rst|abe]
\\
-\sum_x\tfrac12
N\alpha_s\alpha_x\hat{x}\hat{t}\delta_{ab}\delta_{tf}\delta_{e0}\delta(rs;cd;t)
\OLAPtwo(N-1)[xx0|aa0]
\\
-N^2\alpha_r\alpha_s\P(cdf)\theta(cdet)\frac{\hat{f}\hat{t}}{\hat{d}}
\delta_{ds}\usixj ftercd \OLAPtwo(N-1)[rce|abe]
\end{lgathered}
\\
-N^2\alpha_r^2 \hat{r}\hat{e}\delta_{rs}\delta_{ef}\delta_{t0}
\OLAPtwo(N-1)[cde|abe].
\end{multline}
The necessary coupled commutators are given
in~(\ref{eqn-comm-T-A}) and~(\ref{eqn-comm-At-A}).

For $v=3$: The recurrence relation, obtained with
$\Fd(f)\rightarrow\dclusterAC(abeig)$ and
$\Gd(g)\rightarrow\dclusterAC(cdfjh)$, is
\begin{multline}
\label{eqn-recur-Gamma3}
\OBMEthree(N)[cdfjh|rst|abeig]
\\
=\begin{lgathered}[t]
-\sum_x\P(abe)\theta(abghx)\frac{\hat{e}\hat{t}\hat{h}}{\hat{b}\hat{x}}\delta_{bs}
\usixj eightx \usixj etxrab \OLAPthree(N)[cdfjh|raxih]
\\
- \frac{\hat t \hat h}{\hat r}\delta_{is}\usixj eigthr
\OLAPthree(N)[cdfjh|aberh]
-N^2\alpha_s^2\OBMEthree(N-1)[cdfjh|rst|abeig]
\\
+\sum_x\tfrac12N\alpha_s\alpha_x 
\theta(ght)\hat h \hat x \delta_{ft} \delta_{gj}\delta(cd;rs;f)
\OLAPthree(N-1)[xx0jg|abeig]
\\
-\sum_x 
\tfrac12N\alpha_s\alpha_x\P(rst)\P(cdf)\theta(fhrt)\frac{\hat f \hat t \hat x}{\hat
d}\delta_{cg}\delta_{dr}\delta_{js}
\usixj jfhctd  \OLAPthree(N-1)[xx0cg|abeig]
\\
-\sum_x N^2\alpha_r\alpha_s\P(cdf)\theta(cdxt)\frac{\hat f \hat g
\hat t}{\hat d \hat x} \delta_{ds}\usixj fjhgtx \usixj ftxrcd
\OLAPthree(N-1)[rcxjg|abeig]
\\
-N^2\alpha_r\alpha_s\theta(ght)\frac{\hat g \hat t}{\hat
r}\delta_{js} \usixj fjhtgr \OLAPthree(N-1)[cdfrg|abeig]
\end{lgathered}
\\
-N^2\alpha_r^2\hat{r}\hat{g}\delta_{rs}\delta_{gh}\delta_{t0}
\OLAPthree(N-1)[cdfjg|abeig].
\end{multline}
The necessary coupled commutators are given
in~(\ref{eqn-comm-T-AC}) and~(\ref{eqn-comm-At-AC}).

For $v=4$: The recurrence relation, obtained with
$\Fd(f)\rightarrow\dclusterAA(abeijmg)$ and
$\Gd(g)\rightarrow\dclusterAA(cdfklnh)$, is
\begin{multline}
\label{eqn-recur-Gamma4}
\OBMEfour(N)[cdfklnh|rst|abeijmg]
\\
=\begin{lgathered}[t]
\sum_x[1+\theta(emg)\PP{abe}{ijm}]\P(ijm)\theta(ijm)\frac{\hat m \hat
t \hat h}{\hat j \hat x}\delta_{js}\usixj emgthx \usixj mtxrij
\\\qquad\times \OLAPfour(N)[cdfklnh|aberixh]
\\
-N^2\alpha_s^2\OBMEfour(N-1)[cdfklnh|rst|abeijmg]
\\
-\sum_x \tfrac12N\alpha_s\alpha_x [1+\theta(fnh)\PP{cdf}{kln}]
\hat{h}\hat{x}\delta_{fg}\delta_{nt}\delta(kl;rs;n)
\\\qquad\times \OLAPfour(N-1)[xx0cdgg|abeijmg]
\\
+\sum_x \tfrac12N\alpha_s\alpha_x \P(rst)\P(kln)\P(cdf)\hat{f}\hat{n}\hat{g}
\hat{h}\hat{t}\hat{x}\delta_{dr}\delta_{ls}
\ninej tghlkndcf
\\\qquad\times \OLAPfour(N-1)[xx0ckgg|abeijmg]
\\
-\sum_x N^2\alpha_r\alpha_s
[1+\theta(fnh)\PP{cdf}{kln}]\P(kln)\theta(klnght)
\frac{\hat n \hat t \hat g}{\hat l \hat x}\delta_{ls}\usixj fnhtgx
\usixj ntxrkl
\\\qquad\times \OLAPfour(N-1)[cdfrkxg|abeijmg]
\end{lgathered}
\\
-N^2\alpha_r^2\hat{r}\hat{g}\delta_{rs}\delta_{gh}\delta_{t0}
\OLAPfour(N-1)[cdfklng|abeijmg].
\end{multline}
The necessary coupled commutators are  given
in~(\ref{eqn-comm-T-AA}) and~(\ref{eqn-comm-At-AA}).

\section{Recurrence relations for overlaps}
\label{sec-recur-Phi}

The recurrence relations for $\OLAP^{(v)}_N$ are obtained by
rearranging the factors within~(\ref{eqn-overlap-Phi}), as outlined in
the commutation scheme of Sec.~\ref{sec-recur-scheme}.  It is first
necessary to recouple the creation operators within $\Fd(f)$ and
$\Gd(f)$, so as to extract a single-fermion creation operator from
each cluster, giving
\begin{equation}
\label{eqn-subclusters-Phi}
\begin{aligned}
\Fd(f)&=(\Cd(r)\times\Hd(h))^f\\
\Gd(f)&=(\Cd(s)\times\Id(i))^f.
\end{aligned}
\end{equation}
These equations define subclusters $\Hd(h)$ and $\Id(i)$, each
containing $v-1$ fermion creation operators.  Then we deduce the
general recurrence relation
\begin{multline}
\label{eqn-recur-Phi}
\underbracketnote{
\overlap{\S(N)\G(f)}{\S(N)\F(f)}
} {$\OLAP_N^{(v)}[\cdots]$} =
\sum_x (-)^{f-r-h} \hat{f}^{-1}\usixj srxhif
\underbracketnote{
\trme{\S(N)\I(i)}{\T(rsx)}{\S(N)\H(h)}
} {$\OBME_N^{(v-1)}[\cdots]$} 
\\
+\delta_{rs}\delta_{hi}
\underbracketnote{
\toverlap{\S(N)\I(i)}{\S(N)\H(h)}
} {$\OLAP_N^{(v-1)}[\cdots]$}.
\end{multline}
The derivation involves only angular momentum recoupling and the
canonical anticommutator~(\ref{eqn-comm-canonical}).  Each state
appearing in the matrix element and overlap on the right hand side
of~(\ref{eqn-recur-Phi}) has generalized seniority $v-1$.  The overlap
$\OLAP_N^{(v)}$ has therefore been recast in terms of a matrix element
$\OBME_N^{(v-1)}$ and overlap $\OLAP_N^{(v-1)}$, both of
\textit{lower} seniority, providing the basis for recursive calculation with
respect to seniority.

For $v=1$: For this simple special case, with the identifications
$\Fd(f)\rightarrow\dclusterC(c)$ and $\Gd(f)\rightarrow\dclusterC(c)$,
the general recurrence
formula~(\ref{eqn-recur-Phi}) reduces to
\begin{equation}
\label{eqn-recur-Phi1}
\OLAPone(N)[c|c]=\hat{c}^{-1}\OBMEzero(N)[cc0]+\OLAPzero(N).
\end{equation}
Since $\Fd(f)$ and $\Gd(f)$ already consist of single-fermion creation
operators, the ``subcluster creation operators'' $\Hd(h)$ and $\Id(i)$
are simply the identity operator.  From the closed form expression for
$\OBME^{(0)}_N$, this gives
$\OLAPone(N)[c|c]=
\sum_{k=0}^{N-1}(-)^{N+k}(N!^2/k!^2)\alpha_c^{2(N-k)}\OLAPzero(k)+\OLAPzero(N)$.

For $v=2$: The recurrence relation, obtained with
$\Fd(f)\rightarrow\dclusterA(abe)$ and
$\Gd(f)\rightarrow\dclusterA(cde)$, is
\begin{equation}
\label{eqn-recur-Phi2}
\OLAPtwo(N)[cde|abe]
=
\sum_x\theta(abe)\hat{e}^{-1}\usixj{c}{a}{x}{b}{d}{e}\OBMEone(N)[d|acx|b]
+\delta_{ac}\delta_{bd}\OLAPone(N)[b|b].
\end{equation}
The subclusters in~(\ref{eqn-subclusters-Phi}) are $\Hd(h)\rightarrow
\dclusterC(b)$ and $\Id(i)\rightarrow \dclusterC(d)$.
Only the $v=2$ overlaps $\OLAPtwo(N)[aae|aae]$ ($e$ even),
$\OLAPtwo(N)[cc0|aa0]$, and $\OLAPtwo(N)[ace|ace]$ or $\OLAPtwo(N)[cae|ace]$ ($e\neq0$) are
nonvanishing, as may be shown by considering the balance of creation
and annihilation operators in the vacuum expectation value for the
overlap.  Closed form expressions have previously been
obtained~\cite{frank1982:ibm-commutator,lipas1990:ibm-micro-escatt}.

For $v=3$: The recurrence relation, obtained with
$\Fd(f)\rightarrow\dclusterAC(abeig)$ and
$\Gd(f)\rightarrow\dclusterAC(cdfjg)$, is
\begin{multline}
\label{eqn-recur-Phi3}
\OLAPthree(N)[cdfjg|abeig]
\\
=
-\sum_x\theta(fgj)\hat{g}^{-1}\usixj{j}{i}{x}{e}{f}{g}\OBMEtwo(N)[cdf|ijx|abe]
+\delta_{ij}\delta_{ef}\OLAPtwo(N)[cde|abe].
\end{multline}
To decompose $\Fd(f)$ and $\Gd(f)$ according
to~(\ref{eqn-subclusters-Phi}), it is only necessary to reorder the
coupled product so that the single-particle creation operator precedes
the pair creation operator [\textit{i.e.}, $A^\dagger \times C^\dagger
\rightarrow C^\dagger \times A^\dagger$], which then serves as the
subcluster.  Thus, $\Hd(h)\rightarrow (-)^{g-e-i}\dclusterA(abe)$ and
$\Id(i)\rightarrow (-)^{g-f-j}\dclusterA(cdf)$.

For $v=4$: The recurrence relation, obtained with
$\Fd(f)\rightarrow\dclusterAA(abeijmg)$ and
$\Gd(f)\rightarrow\dclusterAA(cdfklng)$, is
\begin{multline}
\label{eqn-recur-Phi4}
\OLAPfour(N)[cdfklng|abeijmg]
\\
= -\sum_{pqx}\hat{g}^{-1}\theta(abdmngq)
\usixj caxpqg \usixj abemgp \usixj cdfngq
\OBMEthree(N)[klndq|acx|ijmbp]
\\
-
\delta_{ac}\sum_p\theta(bdmn)
\usixj abemgp \usixj adfngp
\OLAPthree(N)[klndp|ijmbp].
\end{multline}
Here the clusters must be recoupled as $A^\dagger \times
A^\dagger\rightarrow C^\dagger\times(C^\dagger\times A^\dagger)$.
This gives rise to subclusters $\Hd(h)\rightarrow \sum_p
(-)^{p-b-m}\smallusixj abemgp \dclusterAC(ijmbp)$ and
$\Id(i)\rightarrow \sum_q (-)^{q-d-n}\smallusixj cdfngq
\dclusterAC(klndq)$.

For $v=5$: The recurrence relation, obtained with
$\Fd(f)\rightarrow\dclusterAAC(abeijmgpu)$ and
$\Gd(f)\rightarrow\dclusterAAC(cdfklnhqu)$, is
\begin{multline}
\label{eqn-recur-Phi5}
\OLAPfive(N)[cdfklnhqu|abeijmgpu]
=
\\
-\sum_{x}\hat{u}^{-1}\theta(uqh)
\usixj qpxghu 
\OBMEfour(N)[cdfklmh|pqx|abeijmg]
\\
+
\delta_{pq}\delta_{gh}
\OLAPfour(N)[cdfklmg|abeijmg].
\end{multline}
In this case, since $v$ is odd, the extraction of a single-fermion
creation operator from each cluster requires only
the reordering $(A^\dagger \times A^\dagger)\times C^\dagger
\rightarrow C^\dagger\times(A^\dagger\times A^\dagger)$, 
hence  $\Hd(h)\rightarrow 
(-)^{u-g-p}\dclusterAA(abeijmg)$ and
$\Id(i)\rightarrow (-)^{u-h-q}
\dclusterAA(cdfklmh)$.

\section{Matrix elements of two-body operators}
\label{sec-tbo}

Once overlaps and matrix elements of one-body operators have been
calculated in the generalized seniority scheme, from the recurrence
relations established in Secs.~\ref{sec-recur-Gamma}
and~\ref{sec-recur-Phi}, the matrix elements of two-body operators
follow in a straightforward fashion.    The approach is based, once
again, on expressing the matrix element as a vacuum expectation value,
followed by commutation of the operators to a more convenient
ordering, in which the various terms can be recognized as overlaps and
one-body operator matrix elements.

An arbitrary two-body operator may be decomposed as a linear
combination of elementary terms of the form
$\TBOAA(abecdfw)$.  In particular, if $W^w$ is a spherical tensor
operator of angular momentum $w$, then the angular-momentum coupled
second-quantized form is
\begin{equation}
\label{eqn-tbo-coupled}
W^w=\tfrac14
\sum_{\substack{abcd\\ef}}\deltasqrt{ab}\deltasqrt{cd}\hat{w}^{-1}
\trme{\lclusterA(abe)}{W^w}{\lclusterA(cdf)}_{\text{NAS}} \, \TBOAA(abecdfw),
\end{equation}
where it should be noted that $\TBOAA(abecdfw)=-\TBOCCCC(abecdfw)$,
and where the reduced matrix element is taken with respect to
normalized antisymmetrized states
$\tket{\lclusterA(abe)_\varepsilon}_{\text{NAS}}=\deltasqrtinv{ab}
(\Cd(a)\times\Cd(b))^e_\varepsilon \vket$.  We note this explicitly to
avoid ambiguity with the unnormalized $N=0$, $v=2$ basis state of Sec.~\ref{sec-defn}.  The expression for the
two-body Hamiltonian arises as a special case, with $w=0$.

Evaluating the matrix element of an elementary two-body operator
$\TBOAA(abecdfw)$ as a vacuum expectation value involves only a single
commutation, of the form $[\tilde{A},A^\dagger]\sim T+1$
[see~(\ref{eqn-comm-At-A})]. Schematically,
\begin{multline}
\label{eqn-recur-TBO-schematic}
\trme{\S(N)G}{A^\dagger \tilde{A}}{\S(N)F}
\sim
\vev{(\tilde{G}\St(N))\, (\underbracevalue{A^\dagger \tilde{A}}{T+1}) \, (\Sd(N)F^\dagger)}
\\
\sim
\vev{(\tilde{G}\underbracenovalue{\St(N)\tilde{A}}) \,
 (\underbracenovalue{A^\dagger \Sd(N)}F^\dagger)} +
\vev{(\tilde{G}\St(N))\, T \,(\Sd(N)F^\dagger)}
+
\vev{(\tilde{G}\St(N))\, (\Sd(N)F^\dagger)}
.
\end{multline}
That is, the pair creation and annihilation operators $A^\dagger$ and
$\tilde{A}$, which together constitute the two-body operator, must be
decoupled from each other and reassociated with the fermion cluster
operators for the two states.  If $F$ and $G$ are clusters of
generalized seniority $v$, then the first term above is recognized as
an overlap involving two new clusters, of higher generalized seniority
$v+2$.  
Let us
represent the
reduced matrix elements of the fundamental \textit{two-body} operators
between generalized seniority basis states of \textit{equal}
generalized seniority by 
$\TBME_N^{(v)}[\cdots]$, \textit{e.g.},
\begin{multline}
\label{eqn-defn-TBME}
  \TBMEfour(N)[cdfklnh|rstxyzw|abeijmg] 
\\
\equiv \trme{\S(N)
  \pclusterAA(cdfklnh)}{\TBOAA(rstxyzw)}{\S(N) \pclusterAA(abeijmg)}.
\end{multline} 
Then the resulting expression for the two-body operator matrix
element is of the form
\begin{equation}
\TBMEsymbol{v}{N}\sim \OLAP^{(v+2)}_{N}+\OBME^{(v)}_{N}+\OLAP^{(v)}_{N}.
\end{equation}

The full relation obtained following
this commutation scheme is
\begin{multline}
\label{eqn-recur-TBO}
\trme{\S(N)\G(g)}{\TBOAA(rstxyzw)}{\S(N)\F(f)}=\sum_k(-)^{t+f-k}\hat{k}
\usixj ztwfgk 
\underbracketnote{\overlap{\S(N)I^k}{\S(N) H^k}}{$\OLAP_{N}^{(v+2)}[\cdots]$}\\
+\P(xyz)\P(rst)\theta(sxw) \frac{\hat t \hat z}{\hat r \hat
w}\delta_{ry}
\usixj tzwxsr 
\underbracketnote{\trme{\S(N)G^g}{\T(sxw)}{\S(N)
F^f}}{$\OBME_{N}^{(v)}[\cdots]$}
\\
-{\hat f \hat t} \delta_{tz}\delta_{fg}\delta_{w0} \delta(rs;xy;t)
\underbracketnote{\toverlap{\S(N)G^g}{\S(N) F^f}}{$\OLAP_{N}^{(v)}[\cdots]$},
\end{multline}
in terms
of new clusters defined by
\begin{equation}
\label{eqn-subclusters-TBO}
\begin{aligned}
\Hd(k)&=(\Ad(rst)\times\Fd(f))^k\\
\Id(k)&=(\Ad(xyz)\times\Gd(g))^k.
\end{aligned}
\end{equation}
To obtain a useful relation, in terms of known overlaps of seniority
$v+2$, it may be necessary to recouple the factors
making up the operators $\Hd(k)$ and $\Id(k)$, so that these fermion
cluster creation operators are of the
form used in defining the generalized seniority states in Sec.~\ref{sec-defn}.
For instance, for
the two-body operator matrix elements involving states of $v=3$, 
$\toverlap{\S(N)\I(k)}{\S(N)\H(k)}$ matches the
definition of $\OLAP_N^{(5)}$ in~(\ref{eqn-defn-Phi}) only after 
the recoupling  $[A^\dagger\times(A^\dagger \times
C^\dagger)]\rightarrow [(A^\dagger\times A^\dagger) \times
C^\dagger]$.  

For reference, let us explicitly write the relations for two-body
operator matrix elements for states with generalized seniority $v\leq
3$.  These expressions involve the $\OBME_N^{(v)}$ with $v\leq 4$ and
$\OLAP_N^{(v)}$ with $v\leq 5$, as considered explicitly in
Secs.~\ref{sec-recur-Gamma} and~\ref{sec-recur-Phi}.

For $v=0$:
\begin{multline}
\TBMEzero(N)[rstxyt0]\\
= {\hat t} \OLAPtwo(N)[xyt|rst] 
-\P(xyt)\P(rst)\frac{\hat t}{\hat r}\delta_{rx}\delta_{sy}
\OBMEzero(N)[rr0]
-{\hat t}\delta(rs;xy;t)\OLAPzero(N).
\end{multline}

For $v=1$:
\begin{multline}
\TBMEone(N)[d|rstxyzw|c]\\
=
\begin{lgathered}[t]
-\sum_k\theta(ctk)\hat{k} \usixj ztwcdk \OLAPthree(N)[xyzdk|rstck]
\\
+\P(xyz)\P(rst)\theta(sxw)\frac{\hat t \hat z}{\hat r \hat
w}\delta_{ry}
\usixj tzwxsr \OBMEone(N)[d|sxw|c]
\end{lgathered}
\\
-{\hat c \hat
t}\delta_{tz}\delta_{cd}\delta_{w0}\delta(rs;xy;t)\OLAPone(N)[c|c].
\end{multline}

For $v=2$:
\begin{multline}
\TBMEtwo(N)[cdf|rstxyzw|abe]\\
=
\begin{lgathered}[t]
\sum_k\theta(etk)\hat{k} \usixj ztwefk \OLAPfour(N)[xyzcdfk|rstabek]
\\
+\P(xyz)\P(rst)\theta(sxw)\frac{\hat t \hat z}{\hat r \hat
w}\delta_{ry}
\usixj tzwxsr \OBMEtwo(N)[cdf|sxw|abe]
\end{lgathered}
\\
-{\hat e \hat
t}\delta_{tz}\delta_{ef}\delta_{w0}\delta(rs;xy;t)\OLAPtwo(N)[cde|abe].
\end{multline}

For $v=3$:
\begin{multline}
\TBMEthree(N)[cdfjh|rstxyzw|abeig]\\
=
\begin{lgathered}[t]
-\sum_{pqk}\theta(tgk)\hat{k} \usixj zfqjkh \usixj tepikg \usixj
ztwghk
\OLAPfive(N)[xyzcdfqjk|rstabepik]
\\
+\P(xyz)\P(rst)\theta(sxw)\frac{\hat t \hat z}{\hat r \hat
w}\delta_{ry}
\usixj tzwxsr \OBMEthree(N)[cdfjh|sxw|abeig]
\end{lgathered}
\\
-{\hat g \hat
t}\delta_{tz}\delta_{gh}\delta_{w0}\delta(rs;xy;t)\OLAPthree(N)[cdfjg|abeig].
\end{multline}

\section{Conclusion}
\label{sec-concl}

The calculational framework presented here provides a straightforward
and systematic approach to constructing recurrence relations for
matrix elements and overlaps in a generalized seniority scheme.
Matrix elements of the one-body and two-body multipole operators have
been considered explicitly here, but matrix elements of other
operators of interest, such as the pair transfer
operator~\cite{allaart1988:bpm,barea2009:ibm-doublebeta}, may be
derived similarly.

Aside from some elementary angular-momentum recoupling, the derivation
only requires calculation of commutators of the form
$[T,A^\dagger\times\cdots\times A^\dagger]$ (or
$[T,A^\dagger\times\cdots\times A^\dagger\times C^\dagger]$ for odd
particle number) and $[\tilde{A},A^\dagger\times\cdots\times
A^\dagger]$ (or $[\tilde{A},A^\dagger\times\cdots\times
A^\dagger\times C^\dagger]$ for odd particle number).  This may be
accomplished systematically via the coupled commutator product
rule~(\ref{eqn-comm-prod}).  Although the process becomes increasingly
laborious for larger $v$ (\textit{i.e.}, $v=5$, $6$, and $7$, or higher if
needed), it is also well-suited to automation~\cite{chen1993:wick-clusters}.

\section*{Acknowledgements}
We thank K.~Cai for valuable contributions to the process of
validating the recurrence relations.  Discussions with F.~Iachello,
S.~De Baerdemacker, V.~Hellemans, and P.~Van Isacker are gratefully
acknowledged.  This work was supported by the US DOE under grant
DE-FG02-95ER-40934.

\vfil


\providecommand{\APSLONG}{}
\providecommand{\ELSEVIER}{}
\ELSEVIER\newcommand{\identity}[1]{{#1}}


\end{document}